\documentclass[twocolumn,times]{aastex631}

\usepackage[T1]{fontenc}

\newcommand\ergcms{erg~cm$^{-2}$~s$^{-1}$}
\newcommand\kms{km~s$^{-1}$}

\newcommand{\hbeta}{H{$\beta$}}
\newcommand{\halpha}{H{$\alpha$}}
\newcommand{\OIII}{[O\,{\sc iii}]}
\newcommand{\OIIIa}{[O\,{\sc iii}]\,$\lambda$4959}
\newcommand{\OIIIb}{[O\,{\sc iii}]\,$\lambda$5007}

\newcommand{\HeIIwave}{He\,{\sc ii}\,$\lambda$4686}

\newcommand{\NIIwave}{[N\,{\sc ii}]\,$\lambda$6583}

\newcommand{\SIIa}{[S\,{\sc ii}]\,$\lambda$6716}
\newcommand{\SIIb}{[S\,{\sc ii}]\,$\lambda$6731}

\newcommand{\HeI}{He\,{\sc i}}

\newcommand{\HeIcwave}{He\,{\sc i}\,$\lambda$5876}
\newcommand{\HeIdwave}{He\,{\sc i}\,$\lambda$6678}
\newcommand{\HeIewave}{He\,{\sc i}\,$\lambda$7065}

\shorttitle{A Helium Donor Star in NGC 247 ULX-1}
\shortauthors{Zhou et al.}

\graphicspath{{./}{fig/}}

\begin{document}

\title{Identification of a Helium Donor Star in NGC 247 ULX-1}

\author[0000-0002-5954-2571]{Changxing Zhou}
\affiliation{Department of Engineering Physics, Tsinghua University, Beijing 100084, China}

\author[0000-0001-7584-6236]{Hua Feng}
\affiliation{Department of Astronomy, Tsinghua University, Beijing 100084, China; hfeng@tsinghua.edu.cn}
\affiliation{Department of Engineering Physics, Tsinghua University, Beijing 100084, China}

\author[0000-0002-1620-0897]{Fuyan Bian}
\affiliation{European Southern Observatory, Alonso de C\'{o}rdova 3107, Casilla 19001, Vitacura, Santiago 19, Chile; fuyan.bian@eso.org}

\begin{abstract}
With Very Large Telescope (VLT) Multi Unit Spectroscopic Explorer (MUSE) observations, we detected highly variable helium emission lines from the optical counterpart of the supersoft ultraluminous X-ray source (ULX) NGC 247 ULX-1. No Balmer lines can be seen in the source spectrum. This is the first evidence for the presence of a helium donor star in ULXs, consistent with a prediction that helium donor stars may be popular in ULXs. The helium lines with a FWHM of about 200~\kms\ are likely produced on the outer accretion disk. Their strong variation implies that the central X-ray source can be significantly obscured to the outer disk. Also, a ring or a double-ring structure is revealed in the MUSE image. It is unknown whether or not it is related to the progenitor of the ULX binary. 
\end{abstract}

\section{Introduction}

Ultraluminous X-ray sources (ULXs) are non-nuclear accreting compact objects with an apparent luminosity above the Eddington limit of stellar mass black holes \citep{Fabbiano1989,Makishima2000}. Identifications of high velocity ($\sim$0.1--0.4~$c$) outflows \citep[][]{Pinto2016,Pinto2017,Pinto2021,Walton2016,Kosec2018}, energetic shock-ionized bubble nebulae \citep[][]{Pakull2002,Pakull2003,Ramsey2006,Abolmasov2008,Russell2011,Belfiore2020,Soria2021,Zhou2022}, and neutron star accretors \citep[][]{Bachetti2014,Fuerst2016,Israel2017,Israel2017a,Carpano2018,Sathyaprakash2019,RodriguezCastillo2020}, suggest that the majority of them are powered by supercritical accretion. 

Observations of the optical counterpart of ULXs may shed light on the history of binary evolution \citep{Madhusudhan2008,Patruno2008}, and place constraints on the accretion geometry \citep{Vinokurov2013,Ambrosi2018,Yao2019} and compact object mass \citep{Liu2013,Motch2014}. However, due to their extragalactic distances, spectroscopic observations of their optical counterparts are challenging and only available for a limited number of sources \citep{Abolmasov2007,Roberts2011,Tao2012,Cseh2013,Fabrika2015,Heida2015,Heida2016,Vinokurov2018,Heida2019,Lopez2020}.

Most ULXs are argued to be viewed close to face-on, but there is a class of supersoft ULXs \citep[e.g.,][]{Read2001,Fabbiano2003,Liu2015,Soria2016}, whose X-ray spectrum is dominated by emission below 1~keV, speculated to be viewed close to edge-on \citep[for a review see][]{Kaaret2017}. Thus, these sources could be ideal targets for dynamical measurements using optical spectroscopy. NGC 247 ULX-1 is a typical one in this class \citep{Jin2011}. Mildly relativistic wind with a velocity of 0.17~$c$ is detected in it \citep{Pinto2021}. The source displays frequent dips in the X-ray lightcurve, with transitions between a supersoft regime with little emission above 1~keV and a soft regime with enhanced hard X-ray emission \citep{Feng2016,Alston2021,DA`i2021}. The optical counterpart of NGC 247 ULX-1 was identified by aligning the Chandra and Hubble Space Telescope (HST) images \citep{Tao2012}, has a spectral energy density peaked in the UV band \citep{Feng2016}, which can be reasonably explained as due to self-irradiation from an optically thick wind as a natural consequence of supercritical accretion \citep{Yao2019}. All the above observational features can be interpreted as a result of supercritical accretion viewed at a high inclination angle. In this paper, we present Very Large Telescope (VLT) observations of NGC 247 ULX-1 with the Multi Unit Spectroscopic Explorer (MUSE) instrument \citep{Bacon2010}. 

\begin{figure*}
\centering
\includegraphics[width=0.33\linewidth]{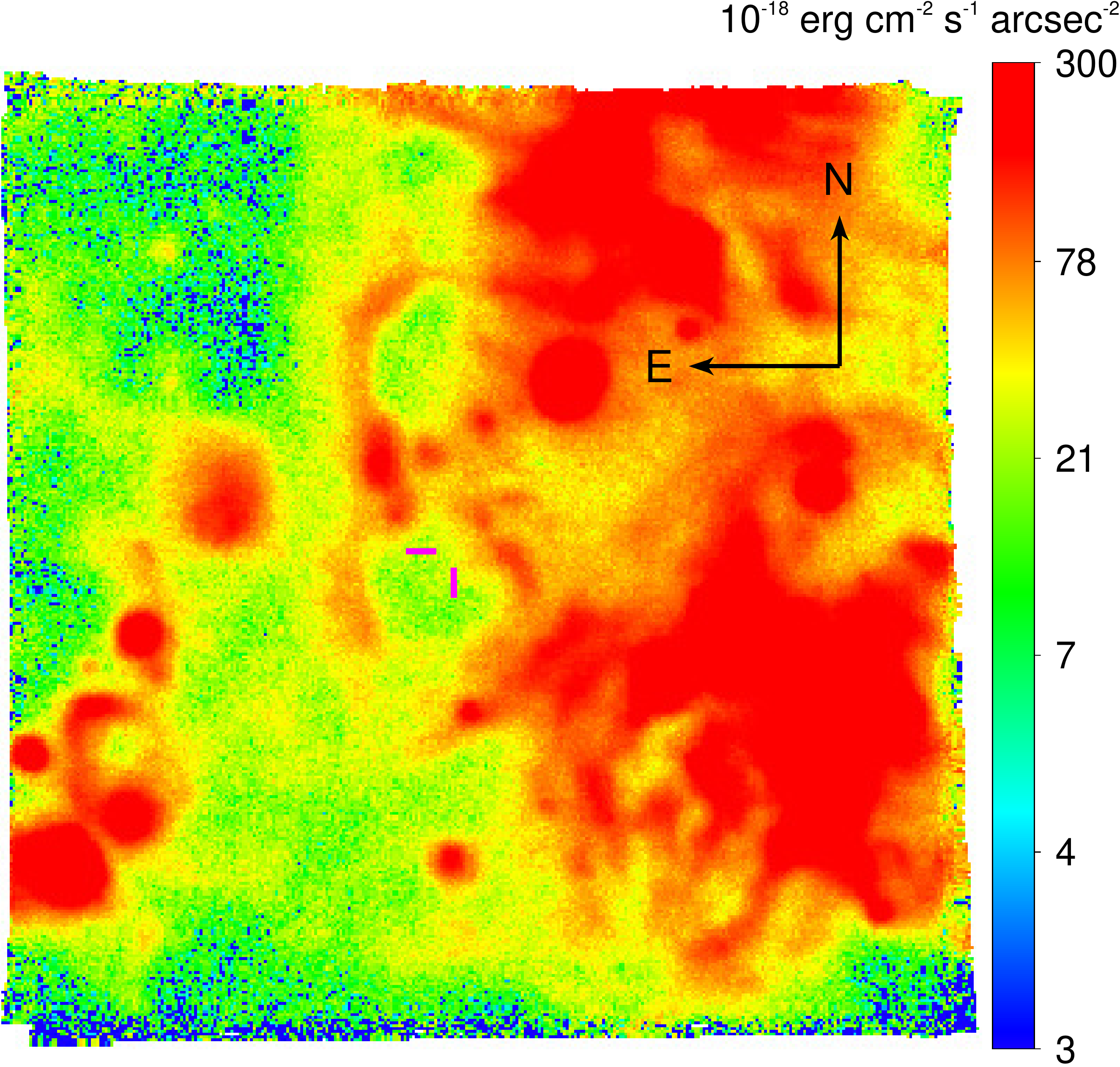}
\includegraphics[width=0.33\linewidth]{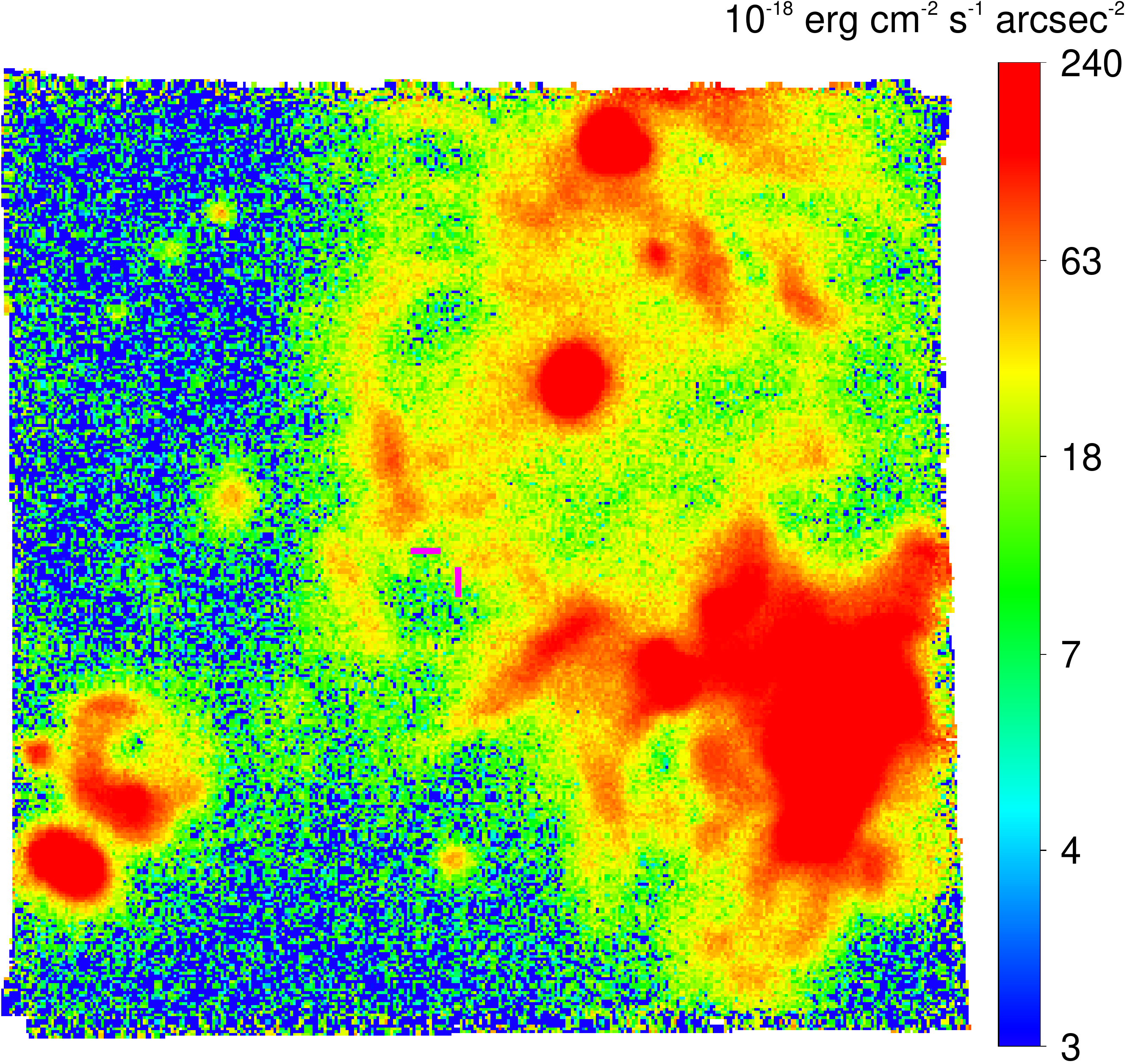}
\includegraphics[width=0.33\linewidth]{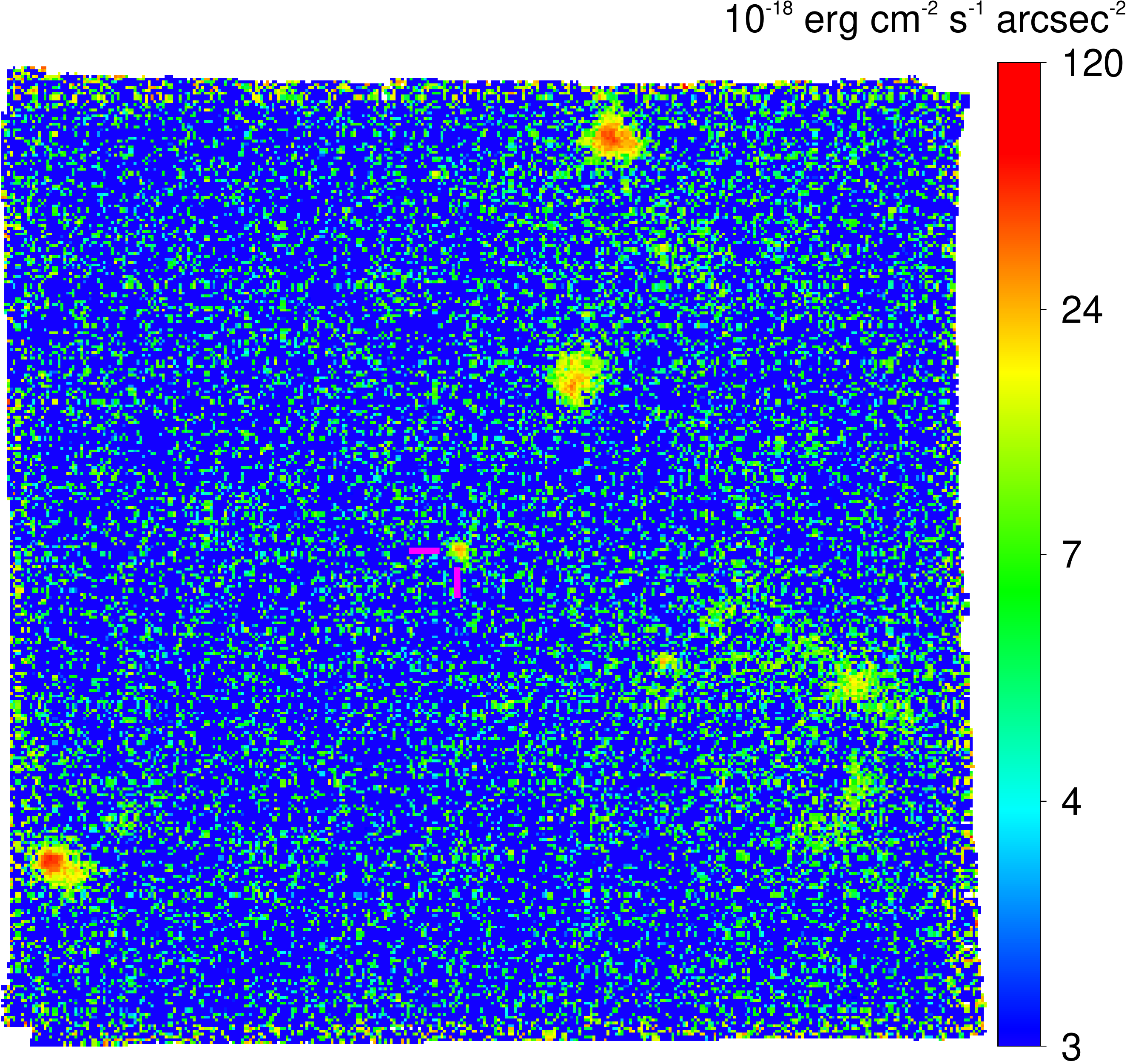}
\caption{MUSE flux images around NGC 247 ULX-1 for \halpha\ (\textbf{left}), \OIIIb\ (\textbf{middle}), and \HeIdwave\ (\textbf{right}), respectively, from combined observations. For \halpha\ and \OIIIb, the flux is extracted from a 10\AA\ window around the line centroid; for \HeIdwave, the extraction window is 11.25\AA. The continuum is estimated from nearby wavelengths free of line features and subtracted. The bars marks the position of the optical counterpart to NGC 247 ULX-1 \citep{Tao2012}. The arrows points north and east, respectively, with a length of 10\arcsec.}
\label{fig:halpha}
\end{figure*}

\begin{figure}
\centering
\includegraphics[width=0.6\columnwidth]{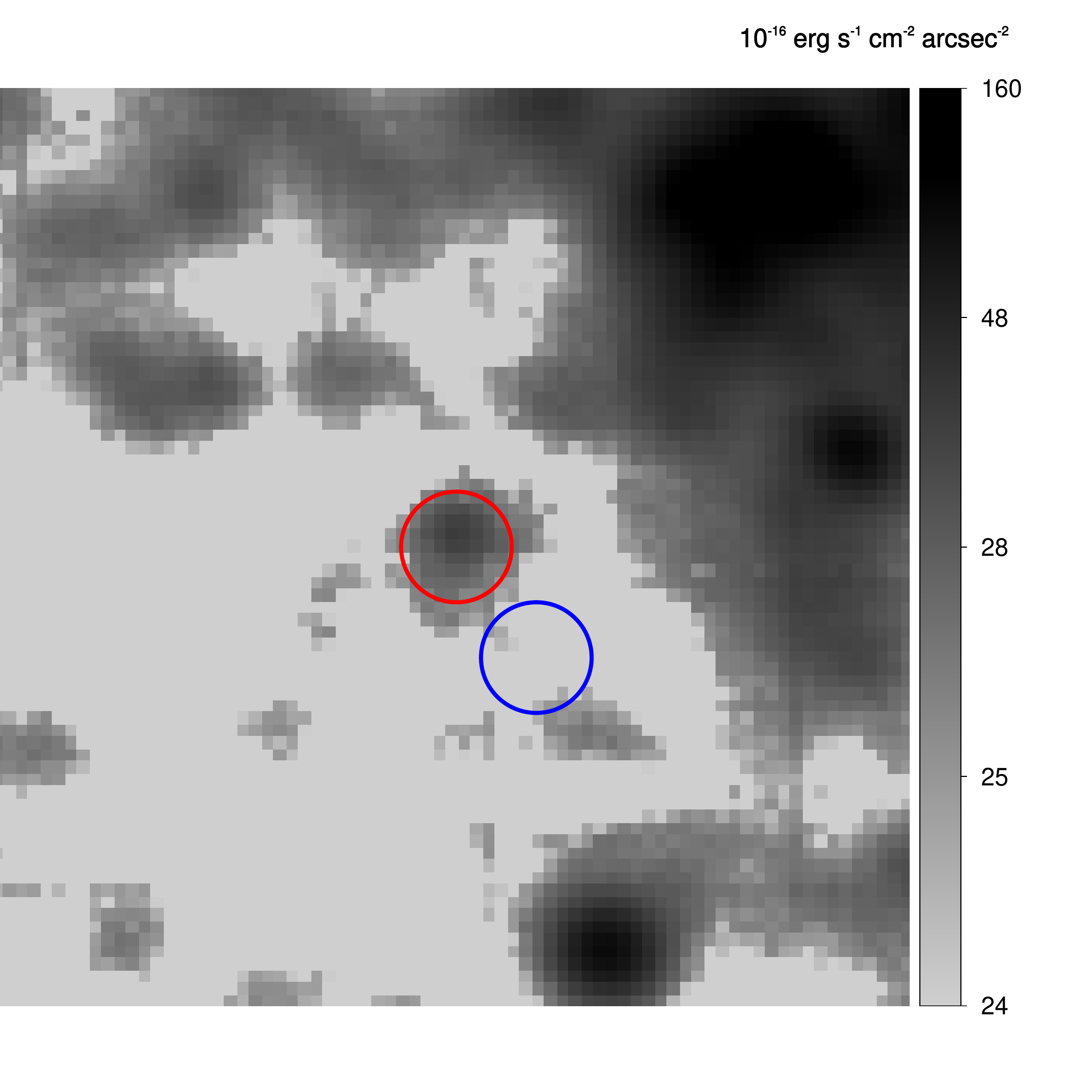}\\
\includegraphics[width=0.6\columnwidth]{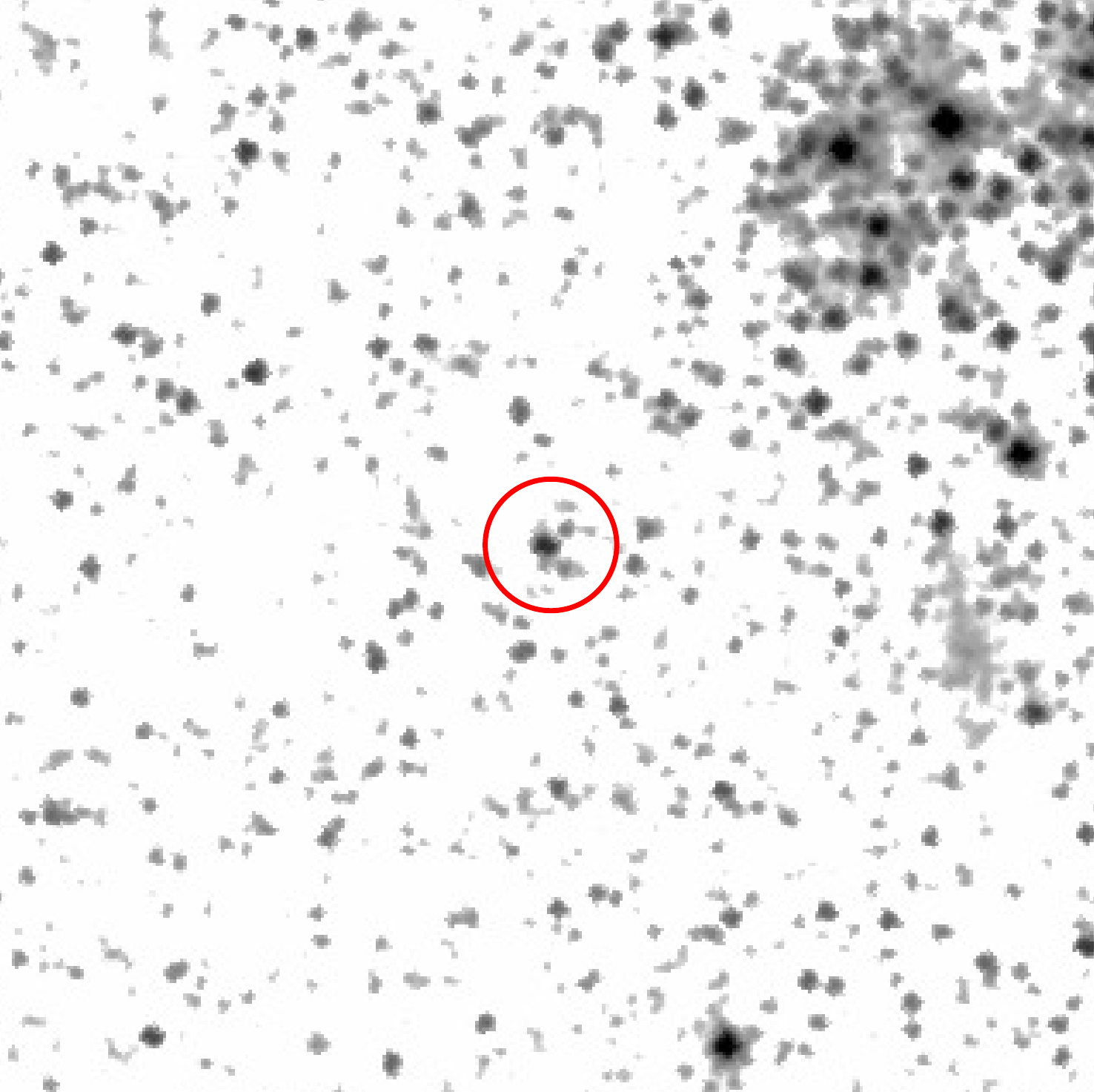}
\caption{{\bf Top}: MUSE image around NGC 247 ULX-1 in the HST ACS WFC F606W filter from combined observations. The red aperture encircles the optical counterpart of the ULX and the blue circle marks the background region, both with a radius of 0\farcs9. {\bf Bottom}: HST ACS WFC F606W image (dataset jblm01030) adopted from \citet{Tao2012}, smoothed with a Gaussian kernel ($\sigma = 1.5$~pixel). The size of the images is 15\arcsec\ $\times$ 15\arcsec.}
\label{fig:f606w}
\end{figure}

\begin{deluxetable}{lccccc}
\tablecaption{Emission line flux, velocity, and velocity dispersion for the optical counterpart of the ULX.}
\label{tab:line}
\tablewidth{\columnwidth}
\tablehead{
\colhead{Line} & \multicolumn{3}{c}{flux} & \colhead{$v$} & \colhead{FWHM} \\
\colhead{} & \multicolumn{3}{c}{($10^{-18}$~\ergcms)}& \colhead{(\kms)} & \colhead{(\kms)} 
}
\startdata
 & May 27 & May 30 & Total & Total & Total \\
\hline
\HeIIwave & $<3 $  & $33 \pm 9$ &  $14 \pm 5$ & $106 \pm 24$ &$120 \pm 106$ \\
\HeIcwave & $39 \pm 5 $  & $50 \pm 5$ &  $42 \pm 4$ & $79 \pm 10$ &$281 \pm 29$ \\
\HeIdwave & $23 \pm 3$  & $32 \pm 3$ &  $26 \pm 2$ & $57 \pm 6$ &$183 \pm16$ \\
\HeIewave & $27 \pm 6$  & $45 \pm 6$ &  $34 \pm 4$ & $57 \pm 9$ &$246 \pm 24$ \\
\enddata
\tablecomments{$v$ is the relative velocity with respect to the host galaxy. FWHM has been corrected for instrument line broadening. Nondetections are quoted as 90\% upper limits. Line fluxes are obtained from the two individual spectra and the total spectrum, respectively, while $v$ and FWHM are obtained from the total spectrum.}
\end{deluxetable}

\begin{figure*}
\centering
\includegraphics[width=\linewidth]{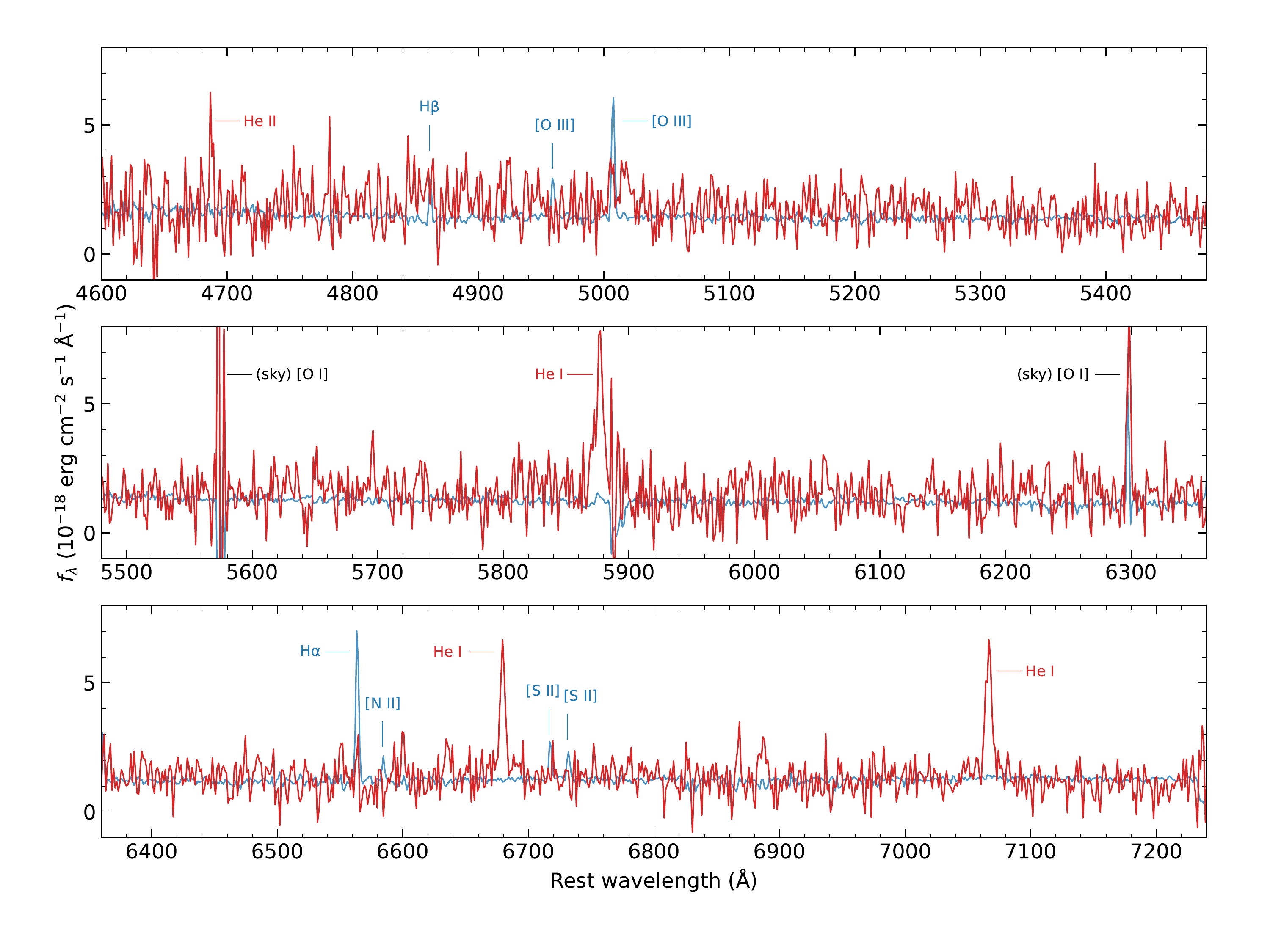}
\caption{MUSE spectrum of the ULX optical counterpart in the wavelength range of 4600--7240 \AA\ from combined observations. The red curve is the background subtracted ULX spectrum, and the blue curve is the background spectrum divided by a factor of 400, extracted from the blue circle region in Figure~\ref{fig:f606w}. Significant emission lines and sky lines are indicated.}
\label{fig:spec}
\end{figure*}

\begin{figure*}
\centering
\includegraphics[width=\linewidth]{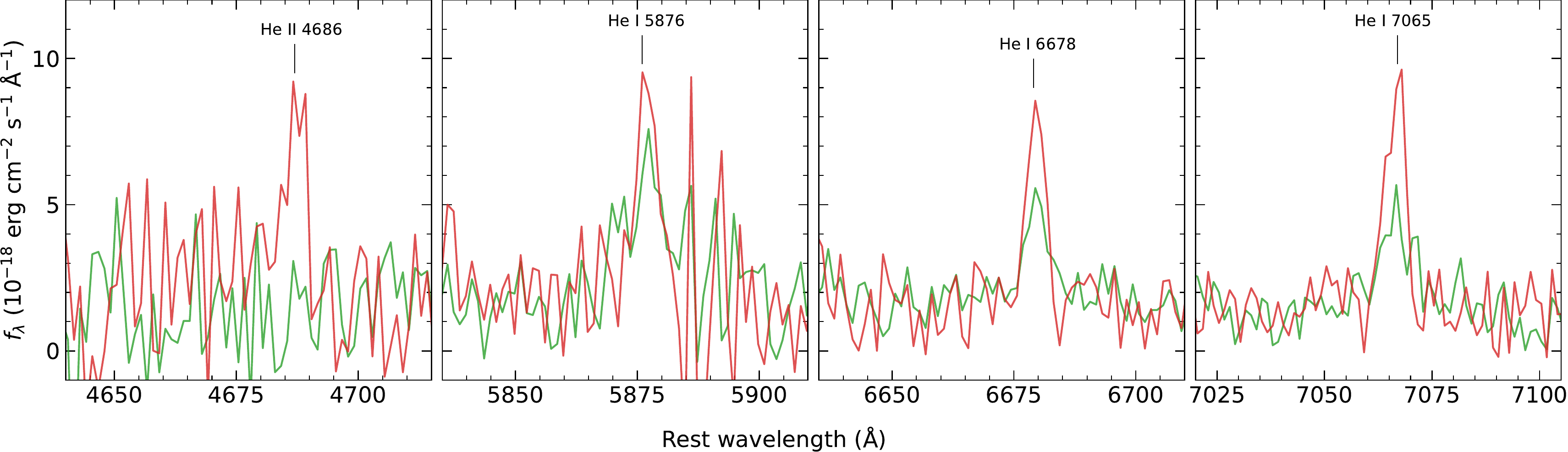}
\caption{Spectra around helium emission lines of the ULX optical counterpart obtained from the two individual observations on May 27 (green) and May 30 (red).}
\label{fig:compare}
\end{figure*}

\section{Observation and data analysis}
\label{sec:obs} 

NGC 247 ULX-1 was observed with VLT MUSE on 2022 May 27 and May 30, respectively, at the La Silla Paranal Observatory under the program 109.238W.002. Each observation contains $4 \times 560$~s on-source exposures and $2 \times 120$~s sky exposures. A 90 degree rotation plus small offset dithering patterns was applied between the on-source exposures. The seeing was 0\farcs5--0\farcs7 during the observations. The sky transparency conditions were clear on the night of May 27, and partially cloudy on the night of May 30. The instrument mode used was WFM-NOAO-E, covering a wavelength range of 4600--9350 \AA.

The data were reduced by the EsoRex pipeline. For each independent exposure, we used the {\it muse\_scibasic} recipe to correct the bias frames, lamp flats, arc lamps, twilight flats, geometry, and illumination exposures. With standard stars taken on the same night, {\it muse\_scipost} was employed to correct the telluric and flux. We used {\it muse\_exp\_align} to align each exposure, and {\it muse\_exp\_combine} to generate the final data cube. The field of view of the final data cube is $1\arcmin \times 1\arcmin$, with the spaxel scale $0\farcs2 \times 0\farcs2$.

Figure~\ref{fig:halpha} shows the MUSE emission line flux images for \halpha, \OIIIb, and \HeIdwave. The flux is summed around the line centroid and the continuum is estimated from nearby wavelengths and subtracted. The position of the optical counterpart to NGC 247 ULX-1 is adopted from \citet{Tao2012} and marked in the plot.  As one can see, there is a point-like source in the ULX region in the \HeIdwave\ image but not in the \halpha\ and \OIIIb\ images. 

To compare with the HST ACS WFC F606W image, we plot the MUSE image extracted in the same filter, shown in Figure~\ref{fig:f606w}. Bright and isolated point-like sources are used to align the two images.  In the MUSE image, there is an object around the ULX position, with a radial profile well consistent with that of nearby point-like sources. The radius that encircles 90\% of the source flux is 0\farcs9, adopted as the source aperture. In the HST image, the optical counterpart of the ULX is the dominant source within the aperture. Thus, the MUSE object in the aperture is mainly attributed to emission from the ULX optical counterpart. 

The absence of the ULX optical counterpart in the \halpha\ image (Figure~\ref{fig:halpha}) suggests that the \halpha\ emission from the source is negligible compared with the diffuse component in the local environment. Therefore, we should choose a background region where the diffuse \halpha\ intensity is similar as in the source aperture. The background region is chosen with the aid of the \halpha\ contour map and shown as a blue circle in Figure~\ref{fig:f606w}. The background subtracted spectrum for the ULX optical counterpart is displayed in Figure~\ref{fig:spec}. 

Several emission lines can be identified from the spectrum. Each emission line is fitted locally with a Gaussian after subtraction of the continuum. The continuum is determined with a linear function from nearby wavelengths free of the emission component. The line flux, velocity relative to the host galaxy \citep[$z = 0.000520$ or $v = 156$~\kms;][]{Koribalski2004}, and the FWHM velocity dispersion corrected for instrument line broadening are listed in Table~\ref{tab:line}. We note that there are a few other helium lines that can be barely seen; we do not report them due to low signal-to-noise ratios. For \halpha, We estimate a 90\% upper limit of $6 \times 10^{-19}$~\ergcms. 

The optical spectrum shows no detectable Balmer lines, consistent with what one can see from the \halpha\ image. The most prominent lines are the \HeI\ lines. \HeIIwave\ can be detected at a significance close to 3$\sigma$. The \OIIIb\ line may arise from the local diffuse component due to incomplete background subtraction, as one can see from Figure~\ref{fig:halpha} that the \OIIIb\ emission in the background region is slightly lower than in the aperture. We checked the background spectrum (also shown in Figure~\ref{fig:spec}) and find no helium emission lines.

The MUSE data contain two individual observations on May 27 and 30, respectively, with a separation of nearly 3~days. To investigate the line variability, we compare the emission line flux obtained from the two individual observations, also listed in Table~\ref{tab:line}. The spectra for several helium lines that show strong variations are displayed in Figure~\ref{fig:compare}. Emission line fluxes from nearby star forming regions are used for relative flux calibration, applied on the spectrum on May 30, with an accuracy estimated to be better than 1\%.  

\begin{figure*}
\centering
\includegraphics[width=\linewidth]{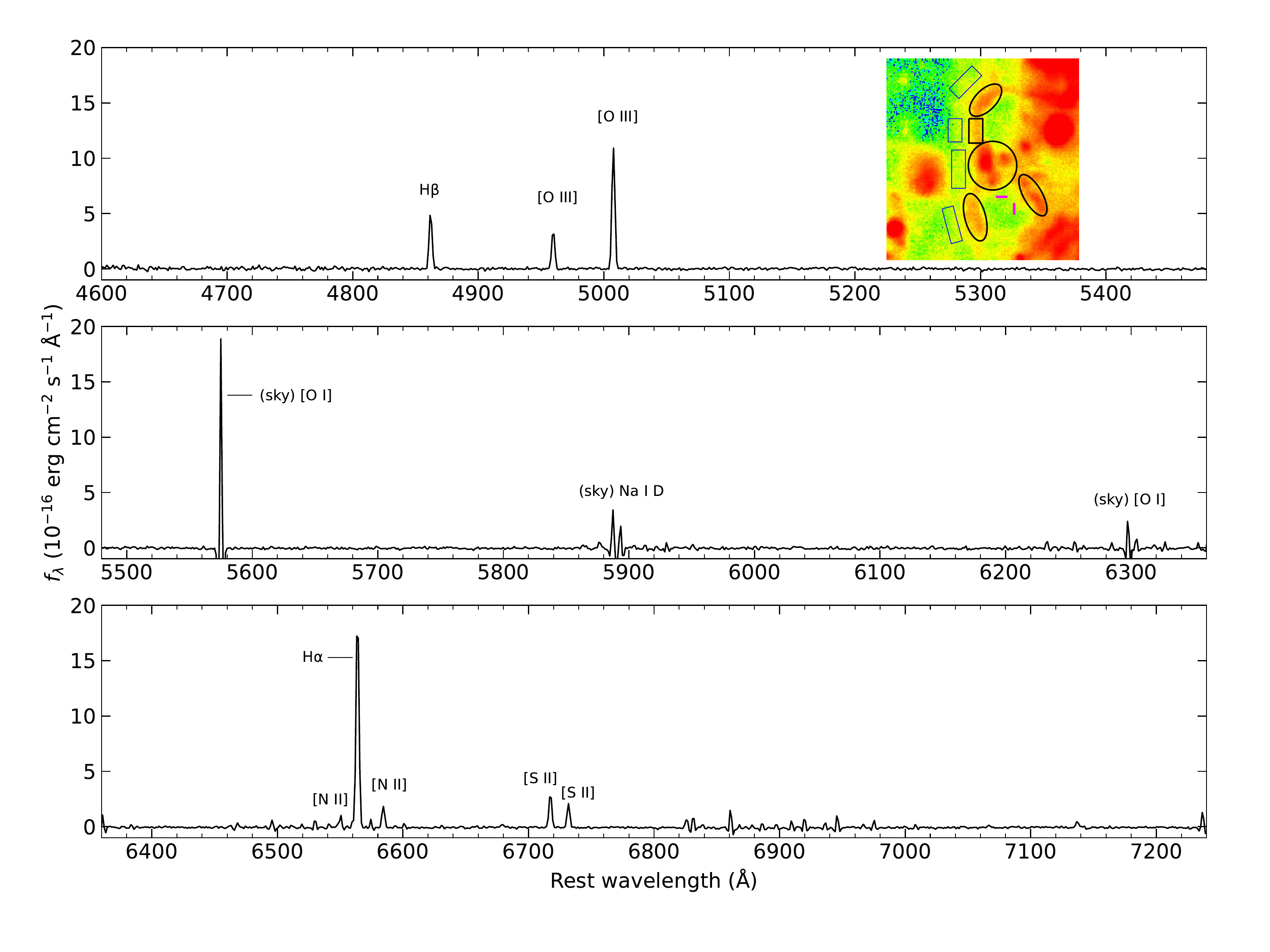}
\caption{Background subtracted spectrum from some parts of the double-ring structure from the combined observation. The inset \halpha\ image shows the extraction regions for the source (thick black) and background (thin blue), and the bars indicate the ULX position. Significant emissions lines are marked.}
\label{fig:ring}
\end{figure*}

Another interesting feature that can be seen in the \halpha\ and \OIII\ images (Figure~\ref{fig:halpha}) around the ULX position is a ring structure, which is possibly part of an ``8'' shaped double-ring structure. The spectrum extracted from some parts of the double-ring structure (to avoid the western star forming region) is displayed in Figure~\ref{fig:ring}. The spectra from different extraction regions look similar, and we thus just display the total spectrum. Each emission line is fitted with a Gaussian, with the line properties listed in Table~\ref{tab:ring}. The intrinsic line width cannot be resolved and is fixed at zero in the fits. For \halpha, the 90\% upper limit of the line FWHM is constrained to be 35~\kms, suggesting that the double-ring structure is due to photoionization instead of shock ionization. The same conclusion can be obtained with the emission line diagnostic diagram \citep{Baldwin1981}.

\begin{deluxetable}{lccl}
\tablecaption{The emission line flux and velocity for the double-ring structure.}
\label{tab:ring}
\tablewidth{\columnwidth}
\tablecolumns{3}
\tablehead{
\colhead{line} & \colhead{flux} & \colhead{$v$} & \\
\colhead{} & \colhead{($10^{-16}$~\ergcms)} & \colhead{(\kms)} 
}
	\startdata
	\hbeta    & $ 15.49 \pm 0.28 $  & $ 41.4 \pm 1.9 $  \\
	\OIIIa    & $ 11.01 \pm 0.14 $  & $ 46.5 \pm 1.2 $  \\
	\OIIIb    & $ 32.81 \pm 0.41 $  & $ 46.5 \pm 1.2 $  \\
	\halpha   & $ 56.02 \pm 0.41 $  & $ 44.6 \pm 0.5 $  \\
	\NIIwave  & $  5.41 \pm 0.14 $  & $ 42.3 \pm 1.9 $  \\ 
	\SIIa     & $  8.74 \pm 0.12 $  & $ 45.0 \pm 0.7 $  \\ 
	\SIIb     & $  5.49 \pm 0.12 $  & $ 45.0 \pm 0.7 $  \\
	\enddata
\tablecomments{$v$ is the relative velocity with respect to the host galaxy. The intrinsic line width is fixed at zero.}
\end{deluxetable}

\section{Discussion}
\label{sec:discuss}

In this work, we detected highly variable helium lines from the optical counterpart of NGC 247 ULX-1 with VLT MUSE observations. The dramatic variation indicates that the emission lines arise from the accretion system. The absence of Balmer lines and richness of helium lines suggests that the mass donor in the X-ray binary is a hydrogen-deficient star. 

\citet{Shao2019} performed a population synthesis study to investigate the neutron star population in ULXs. Combined with detailed calculations of stellar and binary evolutions, they found that in a Milky Way-like galaxy a significant fraction of ULXs are neutron star X-ray binaries containing a helium star companion. The helium star was formed in the common envelope phase during which the hydrogen envelope of its progenitor was stripped \citep[also see][]{Quast2019,Abdusalam2020,Goetberg2020,Wang2021}. Our finding is consistent with their prediction and is the first smoking gun evidence for the presence of a helium donor star in ULXs. Interestingly, \citet{Shao2019} showed that such systems are likely close binaries with an orbital period distribution peaked at 0.1 day, and the helium star mass ranges from 0.6--2~$M_\sun$. 

For NGC 247 ULX-1, no coherent pulsation was detected with a long XMM observation \citep{DA`i2021}. However, this cannot rule out that it contains a neutron star, because the observed X-ray emission may have been largely reprocessed in the wind leading to a low pulsed fraction. Plus, a sufficiently high spin-up rate may be needed for a neutron star ULX to appear as a pulsar \citep{King2017}. Like other soft ULXs, NGC 247 ULX-1 exhibits dips in its lightcurve \citep{Feng2016}. The dips seem to occur randomly in time, and there is no evidence that they are caused by binary motion \citep{DA`i2021,Alston2021}. 

The FWHM of the \HeI\ emission lines is about 180--280~\kms. If we assume a neutron star accretor with a mass of 1.4~$M_\sun$, the FWHM/2 corresponds to a Keplerian velocity at a radius of roughly $(0.5 - 2.0) \times 10^{12}$~cm if the inclination angle is in the range of 45--80 deg (the inclination is supposed to be large but there is no eclipse). Adopting a donor star mass of 0.6--2~$M_\sun$ \citep{Shao2019}, this requires a binary orbital period of at least 2.4--21~day \citep{Eggleton1983} for the Roche-lobe radius around the compact object to be larger than the above radius. This is larger than the most likely orbital period predicted by \citet{Shao2019}, but is consistent with the range of distribution that they predicted. Thus, an outer disk origin for the \HeI\ line is possible in this scenario. Emission lines on the outer accretion disk are supposed to show a double-peak profile, which, however, cannot be detected given the spectral resolution and signal-to-noise ratio of the data. The strong variation of the helium lines implies that the central X-ray source can be largely obscured to the outer accretion disk. Compared with the emission line measured in other ULXs \citep[e.g.,][]{Fabrika2015}, the line width in NGC 247 ULX-1 is much smaller (200 vs.\ 400--1800 \kms), not in support of the scenario that the lines are originating from the base of the accelerating wind as argued for other sources. As per the prediction of \citet{Shao2019} and above estimates, it would be interesting to search for signatures of binary modulation at a period of several days, although the predicted binary period could be as long as 100~days with a low probability. 

%

There is no evidence that the nearby ring or double-ring structure is related to the ULX. One possible conjecture would be that the ring or even the double-ring structure was produced by the progenitor of the compact object in the ULX binary, e.g., similar to one of the scenarios for the formation of the outer rings around the supernova SN~1987A due to magnetically confined anisotropic stellar ejections \citep{Tanaka2002}, although their spatial sizes are distinct.  The ULX binary was kicked off and moved away from the center of the structure to its current position after the supernova explosion. In that case, if we assume that the binary is originally at the center of the southern ring or the conjunction region of the two rings (with a angular distance of 3\arcsec--5\arcsec), a kick-off velocity of at least 5--8~\kms\ can be inferred given a typical age of 10~Myr \citep{Tao2012} and a distance of 3.4~Mpc to the galaxy \citep{Gieren2009}. This is consistent with the velocity difference between the ULX counterpart and ring structure (see Tables~\ref{tab:line} \& \ref{tab:ring}).

\begin{figure}
\centering
\includegraphics[width=\columnwidth]{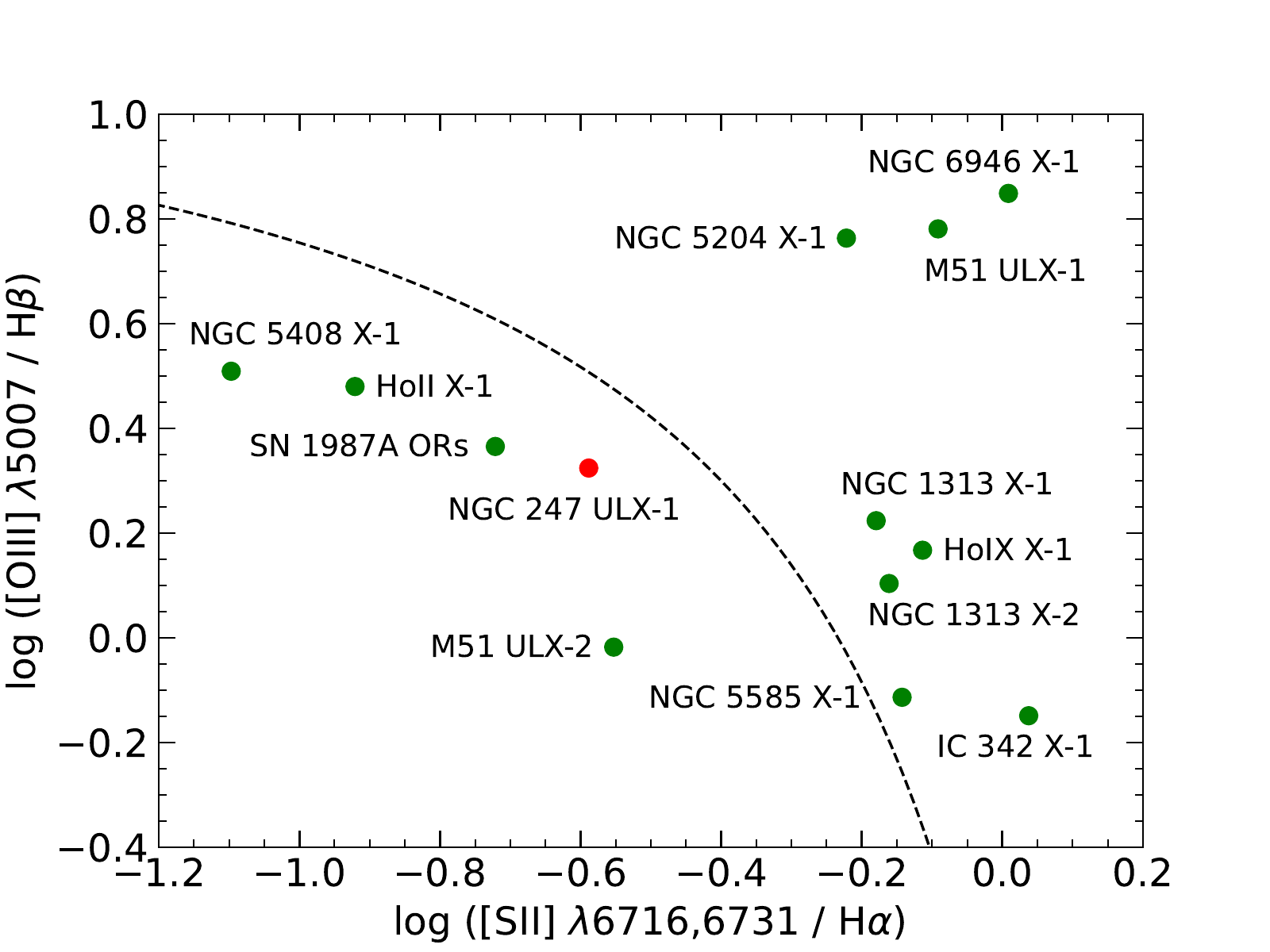}
\caption{The emission line diagnostic diagram \citep{Baldwin1981} for the double-ring structure around NGC 247 ULX-1, ULX bubbles including  
Holmberg II X-1 \citep{Lehmann2005},
Holmberg IX X-1 \citep{Abolmasov2007},
IC 342 X-1 \citep{Roberts2003},
M51 ULX \citep{Urquhart2018},
NGC 1313 X-1 \citep{Gurpide2022} and X-2 \citep{Zhou2022},
NGC 5204 X-1 \citep{Abolmasov2007},
NGC 5408 X-1 \citep{Kaaret2009,Cseh2012},
NGC 5585 X-1 \citep{Soria2021},
NGC 6946 X-1 \citep{Abolmasov2007},
and the SN~1987A outer rings \citep{Tziamtzis2011}.
The dashed curve adopted from \citet{Kewley2001} separates photoionization and shock ionization. }
\label{fig:bpt}
\end{figure}

The small line width and also the line flux ratios suggest that the double-ring structure is a result of photoionization instead of shock ionization. To compare it with other ULX bubbles \citep[e.g.,][]{Pakull2002} as well as the SN~1987A outer rings, we plot their line ratios on the \citet{Baldwin1981} diagnostic diagram, see Figure~\ref{fig:bpt}. It can be seen that the double-ring structure around NGC 247 ULX-1 is similar to the nebula around Holmberg II X-1 \citep{Lehmann2005} and the SN~1987A outer rings \citep{Tziamtzis2011}, both of which are due to photoionization.

\begin{acknowledgments}
We thank the anonymous referee for useful comments. HF acknowledges funding support from the National Key R\&D Project under grant 2018YFA0404502, the National Natural Science Foundation of China under grants Nos.\ 12025301, 12103027, \& 11821303, and the Tsinghua University Initiative Scientific Research Program. Some of the data presented in this paper were obtained from the Mikulski Archive for Space Telescopes (MAST) at the Space Telescope Science Institute. The specific observations analyzed can be accessed via \dataset[DOI: 10.17909/trtf-5s84]{https://doi.org/10.17909/trtf-5s84}.

\end{acknowledgments}


\end{document}